\documentclass{cernyrep}

\newcommand{\cc}{\,\mathrm{cm^{-3}}}
\newcommand{\wcm}{\,\mathrm{W/cm}^2}
\newcommand{\mic}{\,\mu\mathrm{m}}
\newcommand{\mev}{\,\mathrm{MeV}}
\newcommand{\gev}{\,\mathrm{GeV}}
\newcommand{\tev}{\,\mathrm{TeV}}
\newcommand{\tw}{\,\mathrm{TW}}
\newcommand{\gvm}{\,\mathrm{GV/m}}
\newcommand{\mvm}{\,\mathrm{MV/m}}
\newcommand{\tvm}{\,\mathrm{TV/m}}
\newcommand{\fs}{\,\mathrm{fs}}
\newcommand{\ns}{\,\mathrm{ns}}

\newcommand{\mrad}{\,\mathrm{mrad}}

\usepackage{varwidth}
\usepackage{xcolor}

\begin{document}
\title{Plasma Wake Accelerators: Introduction and Historical Overview}
\author{V. Malka}
\institute{%
Laboratoire d'Optique Appliqu\'ee, ENSTA-ParisTech, CNRS, Ecole Polytechnique, Universit\'e Paris-Saclay,  Palaiseau, France\\
}%
\maketitle

\begin{abstract}
Fundamental questions on the nature of matter and energy have found answers thanks to the use of particle accelerators. Societal applications, such as cancer treatment or cancer imaging, illustrate the impact of accelerators in our current life. Today, accelerators use metallic cavities that sustain electric fields with values limited to about 100 ${\mvm}$. Because of their ability to support extreme accelerating gradients, the plasma medium has recently been proposed for future cavity-like accelerating structures. This contribution highlights the tremendous evolution of plasma accelerators driven by either laser or particle beams that allow the production of high quality particle beams with a degree of tunability and a set of parameters that make them very pertinent for many applications.
\end{abstract}

\section{Introduction}

\begin{figure}[ht]
\begin{center}
\includegraphics[width=12cm]{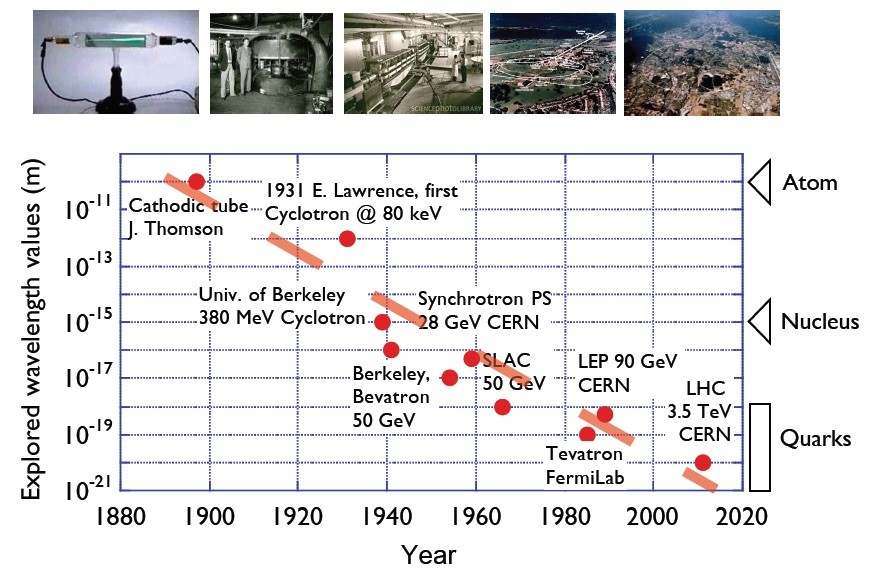}
\caption{\label{cent} Evolution of accelerators and their main related discoveries}
\end{center}
\end{figure}

This article corresponds to the introductory lecture given at the first CAS-CERN Accelerator School on Plasma Wake Acceleration on 21--28 November, 2014. Having this school dedicated to Plasma Accelerators at CERN, where an important part of the story of high energy physics has been written and where the worlds larger accelerators and the brighter and more energetic particle beams are produced, represents in itself the realization of a dream that shows the maturity and the vitality of the field. Having an unexpected level of participation shows also the dynamism of this field of research with an impressive growth of groups in Europe and all over the world. Accelerator physics started almost 130 years ago with the discovery of the cathodic tube. Since then, accelerators have gained in efficiency and in performance delivering energetic particle beams with record energy and luminosity values. During the last century, they have been developed for fundamental research, for example, for producing intense picosecond X-ray pulses in synchrotron machines, or more recently even shorter, few femtosecond X-ray pulses in free electron laser machines. Such short X-ray pulses are crucial for the study of ultra-fast phenomena, for example in biology, to follow the DNA structure evolution, or in material science to follow the evolution of crystals. Higher energy accelerators are crucial to answering important questions regarding the origins of the universe, of dark energy, of the number of space dimensions, etc. The largest one available, the Large Hadron Collider, has for example confirmed two years ago the existence of the Higgs boson. Figure \ref{cent} illustrates few of the many fundamental discoveries that have been made this last century and have permitted matter from $10^{-10}$ to $10^{-20}$ m spatial resolution to be probed.

\begin{figure}[ht]
\begin{center}
\includegraphics[width=12cm]{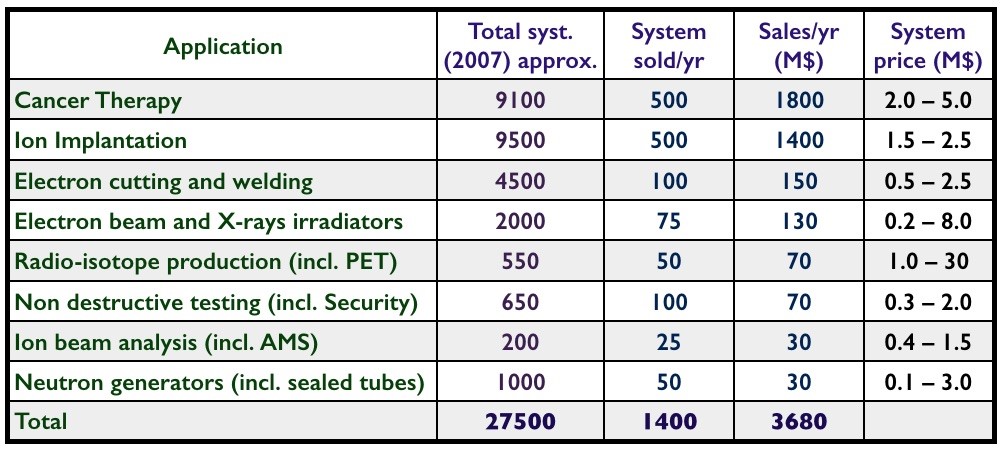}
\caption{\label{market} Market of industrial accelerators and their main societal applications}
\end{center}
\end{figure}

Moreover, as illustrated in Fig.\ref{market}, with an annual market of more than a few US billions of dollars, accelerators are used today in many fields such as cancer therapy, ion implantation, electron cutting and melting, and non-destructive inspection, etc.

\begin{figure}[htbp]
\begin{center}
\includegraphics[width=12cm]{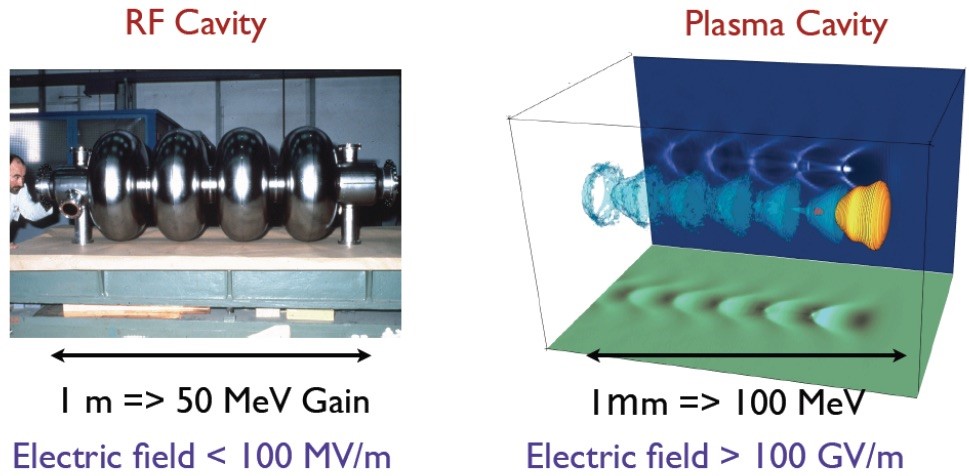}
\end{center}
\caption{Compactness of plasma `cavity'. Left: Radiofrequency cavity. Right: Non-linear laser plasma wakefield. The laser pulse in yellow propagates from left to right, the iso-electronic density is shown in blue and the electron bunch in red.}
\label{cavi}
\end{figure}

The accelerating field in superconducting radio-frequency cavities is, due to electrical breakdown of the metallic cavity, limited to about 100 $\mvm$. It is for this reason that an increase of the particle energy requires an increase of the acceleration length. In the 1950s, Budker and Veksler ~\cite{budk56} proposed using plasma collective fields to accelerate charged particles more compactly. In the pioneering theoretical work performed in 1979, Tajima and Dawson~\cite{taji79} showed how an intense laser pulse can excite a wake of plasma oscillations through the non-linear ponderomotive force associated to the laser pulse. In their proposed scheme, relativistic electrons were injected externally and were accelerated in the very high ${\gvm}$ electric field sustained by relativistic plasma waves. In this early article~\cite{taji79}, the authors proposed two schemes: the laser beat wave and the laser wakefield. Several experiments were performed at the beginning of the 1990s following on from their ideas, and injected few ${\mvm}$ electrons have gained energy in ${\gvm}$ accelerating gradients using either the beat wave or the laser wakefield scheme. In 1994, at the Rutherford Appleton Laboratory, using the 40 ${\tw}$ powerful Vulcan Laser, hundreds of ${\gvm}$ accelerating gradients have been generated and used to trap electrons from the plasma itself, and to accelerate them~\cite{mode95} to few tens of ${\mvm}$ over only 1 mm distance. ${\tvm}$ accelerating gradients have since been demonstrated in the non-linear regime in the forced laser wakefield scheme. Figure \ref{cavi} illustrates the compactness of a plasma accelerating cavity.

In 1985, Chen and Dawson~\cite{chen85} proposed to use a bunched electron beam to drive plasma wakes with, again, ${\gvm}$ accelerating gradients. Soon after, the first experiments on Particle WakeField Acceleration (PWFA) were achieved using low energy electron beam drivers. In 1996, T. Katsouleas and C. Joshi proposed to use an ultra-relativistic electron beam delivered by the SLAC linac to drive ${\gvm}$ accelerating fields. In 2009, the possibility of driving plasma-wakefield acceleration with a proton bunch was proposed~\cite{cald09}, and the authors demonstrated through numerical simulations that ${\tev}$ energy levels could be reached in a single accelerating stage driven by a ${\tev}$ proton bunch.

In both cases, the accelerating gradient results from the rapid electron plasma oscillation that follows the electronic perturbation. If, in an initially uniform and non-collisional plasma a slab of electrons are displaced from their equilibrium position, the restoring force which is applied to this electron slab drives them towards the equilibrium position. For the time scale corresponding to the electrons motion, the motion of the ions can be neglected because of inertia. The typical frequency of electron oscillations around the equilibrium position is called the electron plasma frequency $\omega_\mathrm{pe}$:

\begin{equation}
\omega_\mathrm{pe}=\sqrt{\frac{n_\mathrm{e} e^2}{m_\mathrm{e} \varepsilon_0}}
\label{EqPulsationPlasma},\end{equation}
where $n_\mathrm{e}$ is the unperturbed electron density.

If $\omega_\mathrm{pe}<\omega_0$ (where $\omega_0$ is the laser frequency) then the characteristic time scale of the plasma is longer than the optical period of the incoming radiation. The medium cannot stop the propagation of the electromagnetic wave. The medium is then transparent and it is called `under-dense'. When $\omega_\mathrm{pe}>\omega_0$ then the characteristic time scale of the electrons is fast enough to adapt to the incoming wave and to reflect totally or partially the radiation, and the medium is called `over-dense'.

These two domains are separated at frequency $\omega_0$, which corresponds to the critical density,
$n_\mathrm{c}=~\omega_0^2 m_\mathrm{e} \varepsilon_0 / e^2$. For a wavelength $\lambda_0 = 1 {\mic}$, one obtains $n_\mathrm{c}=1.1 \times 10^{21}{\cc}$. The typical range of electron densities of laser plasma accelerators with current laser technology, is
$[ 10^{17}$ cm$^{-3} - 10^{20}{\cc}]$.

In a uniform ion layer, the density change $\delta n$ for a periodic sinusoidal perturbation of the electron plasma density is written
 \begin{equation}
  \delta n= \delta n_\mathrm{e} \sin(k_\mathrm{p} z-\omega_\mathrm{p} t)
  \label{EqPerturbationDensite},
 \end{equation}
where $\omega_\mathrm{p}$ and $k_\mathrm{p}$ are the angular frequency and the wave number of the plasma wave.

This density change leads to a perturbation of the electric field $\delta \vec{E}$ via the Poisson equation:
 \begin{equation}
  \vec{\nabla}  \delta \vec{E} = - \frac{\delta n \; e}{\varepsilon_0}
    \label{EqMaxwell}.
 \end{equation}

This gives
\begin{equation}
  \delta \vec{E}(z,t)=\frac{\delta n_\mathrm{e} \; e}{k_\mathrm{p} \varepsilon_0} \cos(k_\mathrm{p} z-\omega_\mathrm{p} t) \vec{e_z}.
 \end{equation}

The electric field associated to the relativistic plasma wave, i.e. with a phase velocity close to the speed of light $v_\mathrm{p}=\omega_\mathrm{p}/k_\mathrm{p} \sim c$ can be described by
 \begin{equation}
  \delta \vec{E}(z,t)= E_0 \frac{\delta n_\mathrm{e}}{n_\mathrm{e}} \cos(k_\mathrm{p} z-\omega_\mathrm{p} t) \vec{e_z},
 \end{equation}
where $E_0=m_\mathrm{e} c \omega_\mathrm{pe}/e$.

\begin{figure}[htbp]
   \begin{center}
    \includegraphics[width=10cm]{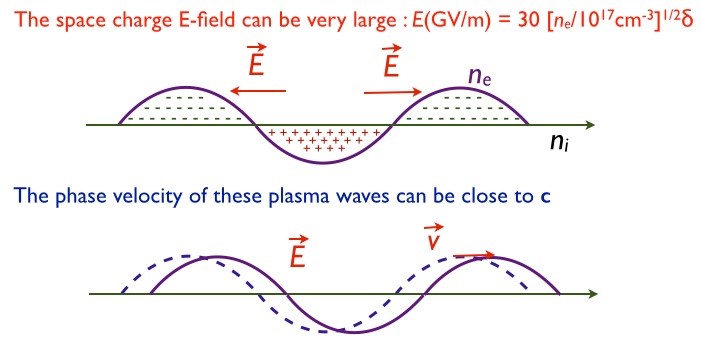}
   \end{center}
   \caption{Density perturbation with the corresponding electric field}
\label{dens}
  \end{figure}

In the linear case, as shown in Fig. \ref{dens}, the relative density perturbation is much smaller than one, and the density perturbation with the electric field has a sinusoidal profile. Note that the electric field is dephased by $- \pi/4$ with respect to the electron density. A $1\%$ density perturbation at a plasma density of $10^{19}{\cc}$ corresponds to 3 ${\gvm}$. In the non-linear case, for a $100\%$ density perturbation at a plasma density of $10^{19}{\cc}$ the accelerating field reaches 300 ${\gvm}$.

\begin{figure}[htbp]
   \begin{center}
    \includegraphics[width=10cm]{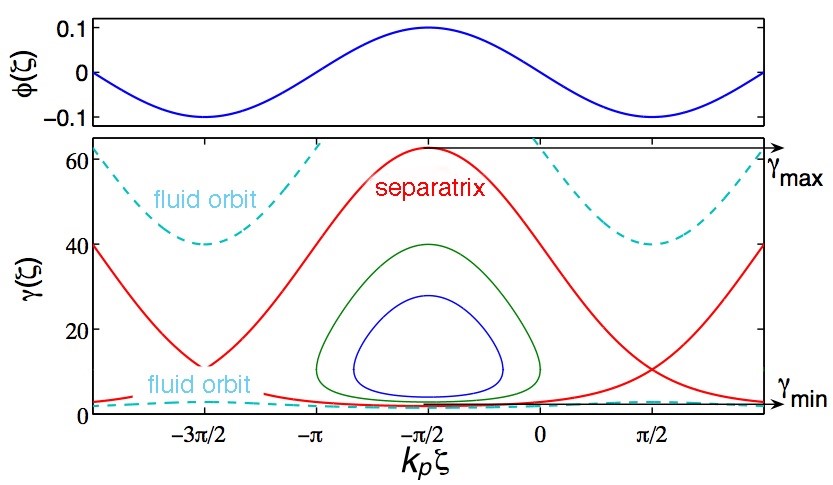}
   \end{center}
   \caption{Upper panel: Potential in phase space. Lower panel: Trajectory of an electron injected ion the potential of the plasma wave in the frame of the wave with the fluid orbit (dashed line), the trapped orbit and in between in red the separatrix.}
 \label{Figfluid_orbit}
  \end{figure}

We now examine the electron motions in this oscillating electric field in the simplified case of a one-dimensional plasma wave. Figure \ref{Figfluid_orbit} represents an example of an electron trajectory in a plasma wave. In this phase space, the closed orbits correspond to trapped particles. Open orbits represent untrapped electrons, either because the initial velocity is too low or to high. The curve which separates these two regions is called the separatrix.
This separatrix gives the minimum and maximum energies for trapped particles. This is comparable to the hydrodynamic case, where a surfer has to crawl to gain velocity and to catch the wave.

  For an electron density much lower than the critical density $n_\mathrm{e} \ll n_\mathrm{c}$, we find $\gamma_\mathrm{p}=\omega_0/\omega_\mathrm{p} \gg 1$ and
  \begin{equation}
   \Delta W_\mathrm{max}= 4 \gamma_\mathrm{p}^2 \frac{\delta n_\mathrm{e}}{n_\mathrm{e}} m c^2
   \label{EqGainMaximalSousDense}.
  \end{equation}

  For an electron travelling along the separatrix, the time necessary to reach maximal energy is infinite because there exists a stationary point at energy $\gamma_\mathrm{p}$. for other closed orbits, the electron successively gains and loses energy during its rotation in the phase space. In order to design an experiment, one needs an estimation of the distance an electron travels before reaching maximal energy gain. This length, which is called the dephasing length $L_\mathrm{deph}$, corresponds to a $\lambda_\mathrm{p}/2$ rotation in the phase space. In order to have a simple analytical estimation, one can assume that the energy gain is small compared to the initial energy of the particle and that the plasma wave is relativistic $\gamma_\mathrm{p} \gg 1$, then the dephasing length is written
  \begin{equation}
   L_\mathrm{deph}\sim\gamma_\mathrm{p}^2 \lambda_\mathrm{p}
   \label{EqLongueurDephasage1D}.
  \end{equation}

  In these formulas, we have considered a unique test electron, which has no influence on the plasma wave. In reality, a massive trapping of particles modifies the electric fields and distorts the plasma wave. This is called the space-charge or beam loading effect (which results from the Coulomb repulsion force). Finally, this linear theory is difficult to apply to non-linear regimes which are explored experimentally.

  \section{Laser wakefield accceleration}

 \subsection{Laser wakefield: the linear regime}
 The ponderomotive force of the laser excites a longitudinal electron plasma wave with a phase equal to the group velocity of the laser close to the speed of light. Two regimes have been proposed to excite a relativistic electron plasma wave.

  In the standard laser wakefield acceleration (LWFA) approach, a single short laser pulse excites the relativistic electron plasma wave. As the ponderomotive force associated with the longitudinal gradient of the laser intensity exerts two
successive pushes in opposite directions on the plasma electrons, the excitation of the electron plasma wave is maximum when the laser pulse duration is of the order of $1/ \omega_\mathrm{p}$.
    For a linearly polarized laser pulse with full width at half maximum (FWHM) $\sqrt{2 \ln 2} \, L$ (in intensity), the normalized vector potential, also called the force parameter of the laser beam, is written

   \begin{equation}
    a(z,t)=a_0 \exp \left[- \left( \frac{k_0 z - \omega_0 t}{\sqrt{2}k_\mathrm{p} L} \right)^2 \right].
   \end{equation}

   In the linear regime, $a_0 \ll 1$, the electronic response obtained behind a Gaussian laser pulse can be easily calculated \cite{gorb87}. In this case, the longitudinal electric field is given by
   \begin{equation}
    \vec{E}(z,t)=E_0 \frac{\sqrt{\pi} a_0^2}{4} k_\mathrm{p} L \exp(-k_\mathrm{p}^2 L^2/4) \cos(k_0 z - \omega_0 t) \vec{e_z}
    \label{EqSillageLaser}.
   \end{equation}

\begin{figure}[htbp]
    \begin{center}
     \includegraphics[width=8cm]{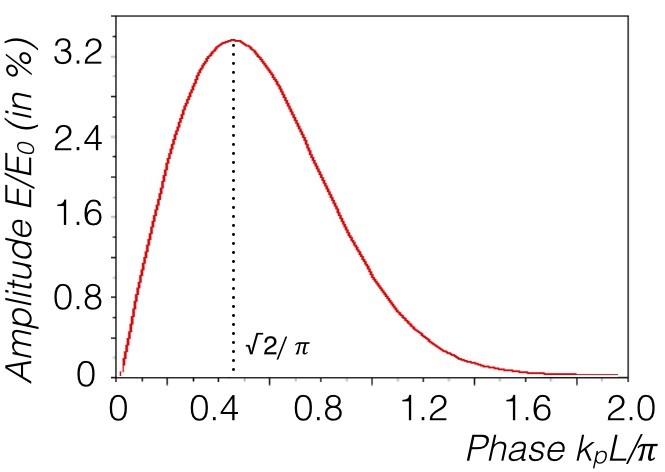}
    \end{center}
    \caption{Amplitude of the electric field as function of the length of a Gaussian laser pulse for a normalized vector potential $a_0=0.3$.}
    \label{FigAmplitudeChpE}
   \end{figure}

\begin{figure}[htbp]
   \begin{center}
    \includegraphics[width=14cm]{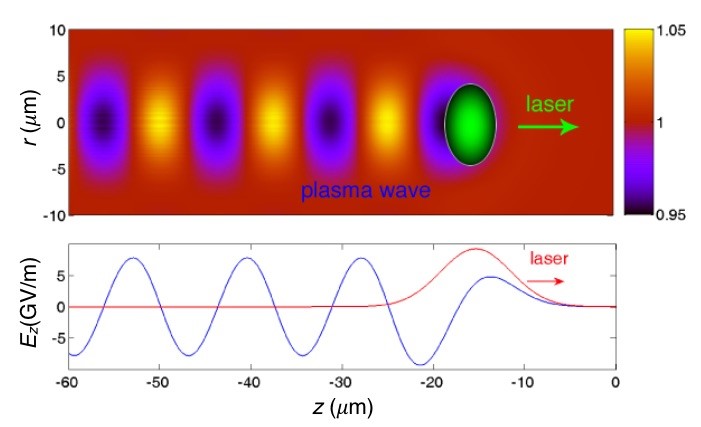}
    \end{center}
   \caption{Density perturbation (top) and electric field (bottom) produced in the linear regime}
    \label{line}
   \end{figure}

   Equation (\ref{EqSillageLaser}) explicitly shows the dependence of the amplitude of the wave on the length of the exciting pulse. In particular, the maximal value for the amplitude is obtained for a length $L=\sqrt{2}/k_\mathrm{p}$ as shown in Fig. \ref{FigAmplitudeChpE} for a laser with a normalized vector potential $a_0=0.3$.
One can note that in the linear regime, the electric field has a sinusoidal shape and reaches maximal values of a few ${\gvm}$. For example, for an electron density $n_\mathrm{e} = 10^{19}$ cm$^{-3}$, the optimal pulse duration equals $L=2.4$ $\mu$m (equivalent to a pulse duration $\tau=8$ fs). For $a_0=0.3$, the maximal electric field is in the ${\gvm}$ range. Figure \ref{line} illustrates the density perturbation and the corresponding longitudinal electric field produced at resonance by a low intensity, $I_\mathrm{laser} = 3 \times 10^{17}$ W$/$cm$^{2}$, laser pulse of 30 s duration.

\begin{figure}[htbp]
    \begin{center}
     \includegraphics[width=8cm]{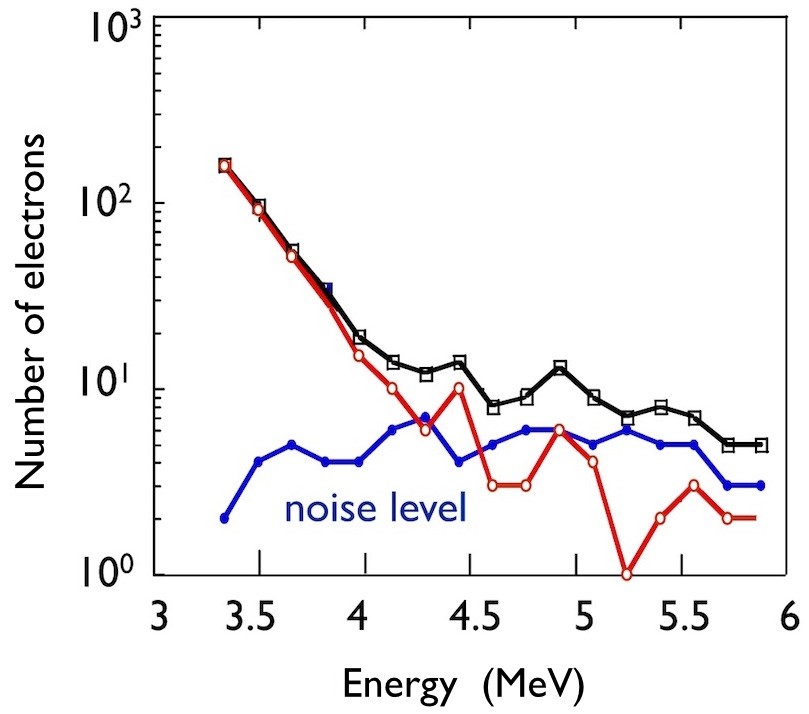}
    \end{center}
    \caption{Electrons spectra obtained at LULI in the laser wakefield scheme}
     \label{sill}
   \end{figure}

In experiments carried out at LULI, relativistic plasma waves with $1\%$ amplitude have been demonstrated. As indicated in Fig. \ref{sill}, 3 ${\mev}$ electrons have been injected into a relativistic plasma wave driven by a 300 fs laser pulse, some of which were accelerated up to 4.6 ${\mev}$\cite{amir98}. The electron spectra has a broad energy distribution with a Maxwellian like shape, as expected when injecting an electron beam with a duration much longer that the plasma period, and in this case with a duration much longer than the plasma wave live-time.

 \subsection{Laser beatwave}

\begin{figure}[htbp]
    \begin{center}
     \includegraphics[width=8cm]{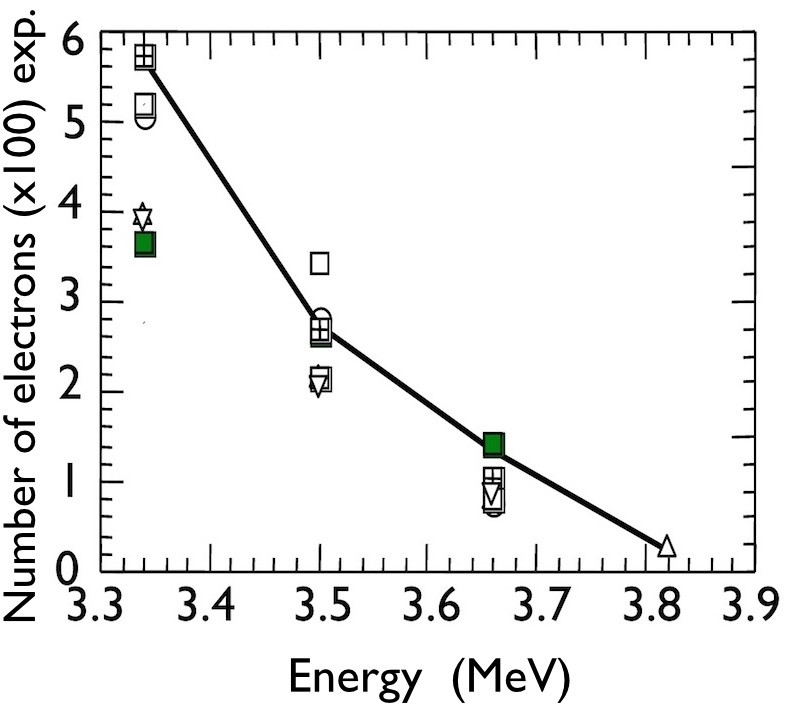}
    \end{center}
    \caption{Electrons spectra obtained at LULI in the laser beat wave scheme}
     \label{FigLULI}
   \end{figure}

 Before the advent of short and intense laser pulses, relativistic plasma waves were driven by the beatwave of two long laser pulses of a few tens of picoseconds (i.e. with duration much greater than the plasma period). In this case, the plasma frequency $\omega_\mathrm{p}$ has to satisfy exactly the matching condition, $\omega_\mathrm{p}~=~\omega_{1}-~\omega_{2}$, with $\omega_{1}$ and $\omega_{2}$ the frequencies of the two laser pulses.
 The first observation of relativistic plasma waves was performed using the Thomson scattering technique by the group of C. Joshi at UCLA~ \cite{clay85}. Acceleration of 2 ${\mev}$ injected electrons up to 9 ${\mev}$ \cite{clay93} and later on, up to 30 ${\mev}$~ \cite{ever94}, were demonstrated by the same group using a CO$_{2}$ laser of about 10 $\mu$m wavelength. At LULI, 3 ${\mev}$ electrons were accelerated up to 3.7 ${\mev}$ in beat wave experiments with Nd:Glass lasers of about 1 $\mu$m wavelengths by a longitudinal electric field of 0.6 ${\gvm}$ ~\cite{amir95}. Similar works were also performed in Japan at University of Osaka ~\cite{kita92}, in the UK at the Rutherford Appleton Laboratory ~\cite{dyso96}, and in Canada at the Chalk River Laboratory~\cite{ebra94}. Electron spectra obtained at LULI in the laser beat wave scheme are shown on Fig.~\ref{FigLULI}.

In order to reduce the coupling between electron waves
and ion waves which was a limiting factor of previous experiments performed
with 100 ps Nd lasers \cite{amir92}, experiments done at
the Rutherford Appleton Laboratory with a 3 ps laser pulse have shown excitations of
higher amplitude relativistic plasma waves \cite{walt02}.

 \subsection{Self-modulated laser wakefield}
In all of these experiments, because of the duration of the injected electron bunch, which is much longer than the plasma period and even longer than the life time of the plasma, only a very small fraction of injected electrons were accelerated and the output beam had a very poor quality with a Maxwellian-like energy distribution.

\begin{figure}[ht]
   \begin{center}
    \includegraphics[width=14cm]{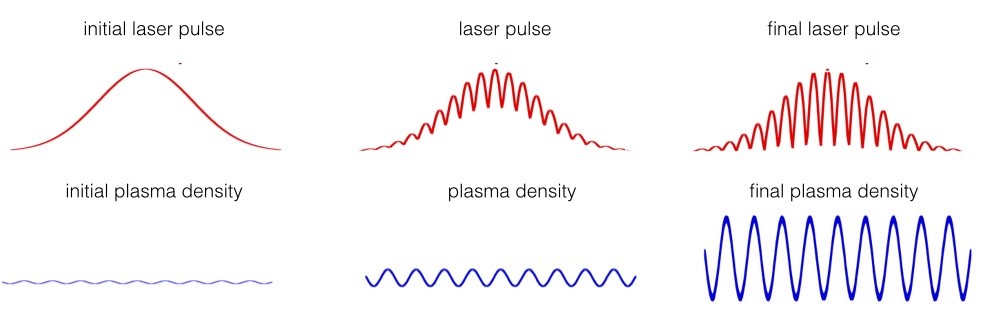}
    \end{center}
   \caption{Evolution of the laser pulse and plasma density in the self-modulated laser wakefield regime}
\label{autom}
     \end{figure}

 Thanks to the development of powerful laser systems with short pulse duration (500 fs), a new regime that allows self-injection of electrons in very intense accelerating gradients with values exceeding 100${\gvm}$ has been discovered. The cumulative effects of the self-focusing and the self-modulation of the laser envelope by the initial perturbation of the electron plasma density generates a train of laser pulses which become resonant with the plasma wave. These effects are described in Fig. \ref{autom}. The self-modulated laser wakefield regime occurs when the laser pulse duration exceeds the plasma period and when the laser power exceeds the critical power for self-focusing ~\cite{spra92b,anto92,andr92}. The initial Gaussian laser pulse becomes modulated at the plasma wavelength during its propagation. This mechanism, which is close to a Forward Raman Scattering Instability ~\cite{josh81}, can be described as the decomposition of an electromagnetic wave into a plasma wave at a frequency shifted by the plasma frequency.

In an experiment done at the Rutherford Appleton Laboratory, a relativistic plasma wave was excited by an intense laser (${>} 5\times 10^{18}$ W$/$cm$^2$), for a short duration (${<}1$ ps), by a 1.054 $\mu$m wavelength laser pulse in the self-modulated laser wakefield regime. This is the decay (induced by a noise level plasma wave) of the strong electromagnetic pump wave ($\omega_{0}, k_{0}$) into the plasma wave ($\omega_\mathrm{p}, k_\mathrm{p}$) and two forward propagating electromagnetic cascades at the Stokes ($\omega_{0}-n\omega_\mathrm{p}$) and anti-Stokes ($\omega_{0}+n\omega_\mathrm{p}$) frequencies, $n$ being a positive integer, and $\omega$ and $k$ being the angular frequency and the wavenumber, respectively, of the indicated waves. The spatial and temporal interference of these sidebands with the laser produces an electromagnetic beat pattern propagating synchronously with the plasma wave. The electromagnetic beat exerts a force on the plasma electrons, reinforcing the original noise level plasma wave which scatters more sidebands, thus closing the feedback loop for the instability.

\begin{figure}[ht]
   \begin{center}
    \includegraphics[width=16cm]{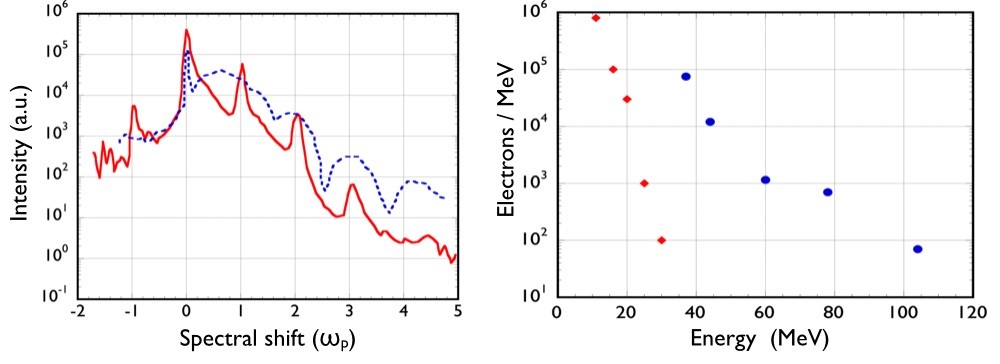}
    \end{center}
   \caption{Frequency and electron spectrum the self-modulated laser wakefield regime for two different electron plasma densities: $ 0.54\times 10^{18}$ cm$^{-3}$ (in red) and $ 1.5\times 10^{19}$ cm$^{-3}$ (in blue).}
\label{ral}
     \end{figure}

The solid curve in Fig. \ref{ral} shows the electromagnetic frequency spectrum emerging form the plasma with a density of ${>}5\times 10^{18}$ cm$^{-3}$, where the abscissa is the shift in frequency of the forward scattered light from the laser frequency in units of $\omega_\mathrm{p}$. The upshifted anti-Stokes and downshifted Stokes signals at $\Delta\omega/\omega_\mathrm{p}=\pm1$ are clearly visible as is the transmitted pump at $\Delta\omega/\omega_\mathrm{p}=0$ and the second and third anti-Stokes sidebands. These signals are sharply peaked, and their widths indicate that the plasma wave which generated these signals must have a coherence time of the order of the laser pulse. The dashed curve shows the spectrum when the density is increased to $ 1.5\times 10^{19}$ cm$^{-3}$. The most startling feature is the tremendous broadening of the individual anti-Stokes peaks at this higher density. This broadening corresponds to wave-breaking and is mainly caused by the loss of coherence due to severe amplitude and phase modulation as the wave breaks. As wave-breaking evolves, the laser light no longer scatters off a collective mode of the plasma but instead scatters off the trapped electrons which are still periodically deployed in space but have a range of momenta producing, therefore, a range of scatter frequencies.

   During experiments carried out in the UK in 1994 \cite{mode95}, the amplitude of the plasma waves reached the wave-breaking limit, where electrons initially belonging to the plasma wave are self-trapped and accelerated to high energies. The fact that the external injection of electrons in the wave is no longer necessary is a major improvement. Electron spectrums extending up to 44 ${\mev}$ have been measured during this first campaign, and up to 104 ${\mev}$ in the second campaign. This regime has also been reached for instance in the United States at CUOS \cite{umst96}, and at NRL \cite{moor97}. However, because of the heating of the plasma by these relatively `long' pulses, the wave-breaking occurred well before reaching the cold wave-breaking limit, which limited the maximum electric field to a few 100 ${\gvm}$. The maximum amplitude of the plasma wave has also been measured to be in the range 20--60\% \cite{clay98}.

\begin{figure}[htbp]
   \begin{center}
    \includegraphics[width=14cm]{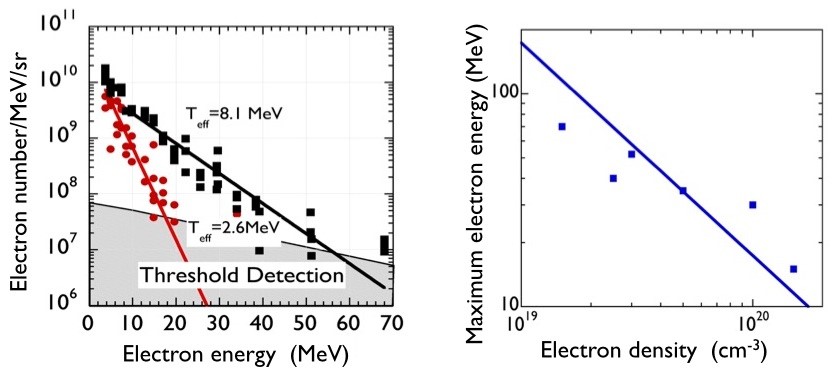}
    \end{center}
   \caption{Left: Typical electron spectra obtained at $ 5\times 10^{19}$ cm$^{-3}$ (squares) and $ 1.5\times 10^{20}$ cm$^{-3}$ (circles). The corresponding effective temperatures are 8.1 ${\mev}$ (2.6  ${\mev}$) for electron density of $ 5\times~10^{19}$~cm$^{-3}$~($ 5\times 10^{20}$~cm$^{-3}$). Right: Maximum electron energy as a function of the plasma electron density. Experimental data: squares. Theoretical data: line.}
    \label{sloa2001}
   \end{figure}

 Experiments performed at LOA since 1999 have shown that an electron beam can also be produced using a compact 10 Hz laser system \cite{malk01}. Fig. ~\ref{sloa2001} shows two typical electron spectra obtained at $ 5\times 10^{19}$~cm$^{-3}$ and $ 1.5\times 10^{20}$ cm$^{-3}$. The 0.6 J, 35 fs laser beam was focused tightly to a 6 $\mu$m focal spot leading to a peak laser intensity of $ 2\times 10^{19}$ W$/$cm$^{2}$. Electron distributions with electron energy greater than 4 ${\mev}$ are well fitted by an exponential function, characteristic of an effective temperature for the electron beam. These effective temperatures are 8.1 ${\mev}$ (2.6 ${\mev}$) for electron density of $ 5\times 10^{19}$~cm$^{-3}$ ($1.5\times 10^{20}$ cm$^{-3}$), to which correspond typical values of 54 ${\mev}$ (20 ${\mev}$) for the maximum electron energy. This maximum energy is defined by the intersection between the exponential fit and the detection threshold. One can observe an important decrease in the effective temperature and in the maximum electron energy for increasing electron densities.

	This point is summarized in Fig.~\ref{sloa2001} where we present the maximum electron energy as a function of the electron density. It decreases from 70 ${\mev}$ to 15 ${\mev}$ when the electron density increases from $ 1.5\times 10^{19}$ cm$^{-3}$ to $ 5\times 10^{20}$ cm$^{-3}$. Also presented in Fig.~\ref{sloa2001} is the theoretical value \cite{esar96}

     \begin{equation}
W_\mathrm{max} \approx 4\gamma_\mathrm{p}^2 (E_z/E_0)mc^2 F_\mathrm{NL}.
\end{equation}

Here, the maximum electron energy is greater than the conventional one given by the simple formula $W_\mathrm{max} \approx 2\gamma_\mathrm{p}^2 (E_z/E_0) mc^2$, where $\gamma_\mathrm{p}$ is the plasma wave Lorentz factor (which is equal to the critical density to electron density ratio $n_\mathrm{c}/n_\mathrm{e}$) and $E_z/E_0$ is the electrostatic field normalized to $E_0$ ($E_0=cm\omega_\mathrm{p}/e$). The factor of two is due to self-channelling induced by the space-charge field which focuses accelerated electrons for all phases. The correction factor $F_{NL}\approx(\gamma_{\perp0}n_0/n)^{3/2}$ corresponds to a non-linear correction due to the relativistic pump effect and to self-channelling. In this formula, $n_0$ is the initial electron density, $n$ the effective one and $\gamma_{\perp0}$ is the Lorentz factor associated to the laser intensity: $\gamma_{\perp0} =(1+a_0^2/2)^{1/2}$. The electron density depression is estimated by balancing the space-charge force and laser ponderomotive force, and evaluated by $\delta n/n = (a_0^2/2\pi^2)(1+a_0^2/2)^{-1/2}(\lambda_p/w_0)^2$.

In the lower electron density case, the depression correction will introduce an important increase of the maximum energy gain which is multiplied by a factor of 2 at $ 1.5\times 10^{19}$ cm$^{-3}$. For densities greater than $ 1.0\times 10^{20}$ cm$^{-3}$, the main contribution is due to the relativistic pump effect, as outlined on the plot in Fig. \ref{sloa2001}. It is also crucial to note that the fact that the electron maximum energy increases when the electron density decreases demonstrates that electrons are mainly accelerated by relativistic plasma waves. The maximum electron energy calculated at lower density overestimated the experimental ones, indicating that the dephasing length becomes shorter than the Rayleigh length. In order to solve this problem, experiments were performed at LOA using a longer off-axis parabola, more energetic electrons have been measured, with a peak laser intensity ten times smaller than in this first experiment.

Electron beams with Maxwellian spectral distributions, generated by compact high repetition rate ultra-short laser pulses, have been also at this time been produced in many laboratories around the world: at LBNL \cite{leem04}, at NERL \cite{hoso03} and in Germany \cite{gahn99} for instance, and are now currently produced in more than 20 laboratories worldwide.

  \subsubsection{Forced laser wakefield} \label{SecSillageForce}
\begin{figure}[htbp]
   \begin{center}
    \includegraphics[width=14cm]{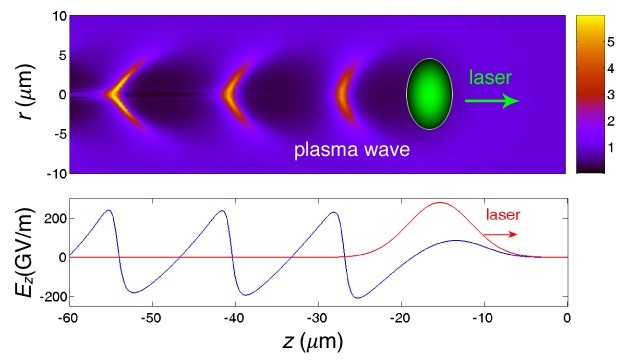}
    \end{center}
   \caption{Density perturbation (top) and electric field (bottom) produced in the non-linear regime}
 \label{Figpp}
     \end{figure}

The forced laser wakefield regime \cite{malk02} is reached when the laser pulse duration is approximately equal to the plasma period and when the laser waist is about the plasma wavelength. This regime allows a reduction in heating effects that are produced when the laser pulse interacts with trapped electrons. In this regime, highly non-linear plasma waves can be reached as can be seen in Fig.~\ref{Figpp}.

   The laser power needs also to be greater that the critical power for relativistic self-focusing in order for the laser beam to shrink in time and in space. Due to self-focusing, pulse erosion can take place, which can allow efficient wake generation. Since the very front of the pulse is not self-focused, the erosion will be more severe. The wake then is mostly formed by this fast rising edge, and the back of the pulse has little interaction with the relativistic longitudinal oscillation of the plasma wave electrons. Indeed, the increase of plasma wavelength due to relativistic effects means that the breaking and accelerating peak of the plasma wave sits behind most, if not all, of the laser pulse. Hence its interaction, and that of the accelerated electrons with the laser pulse, is minimized, thus reducing emittance growth due to direct laser acceleration. Thanks to short laser pulses, plasma heating in the forced laser wakefield regime is significantly lower than in the self-modulated wakefield regime. This allows much higher plasma wave amplitudes to be reached, as well as higher electron energies. Thanks to a limited interaction between the laser and the accelerated electrons, the quality of the electron beam is also improved. Indeed, the normalized transverse emittance measured using the pepper pot technique has given values comparable to those obtained with conventional accelerators with an equivalent energy (normalized r.m.s. emittance $\varepsilon_n = 3 \pi$ mm mrad for electrons at $55 \pm 2$ ${\mev}$) \cite{frit04}.

The three-dimensional simulations realized for this experiment showed that the radial plasma wave oscillations interact coherently with the longitudinal
field, so enhancing the peak amplitude of the plasma wave. This, coupled with the aforementioned strong self-focusing,
are ingredients absent from one-dimensional treatments of this interaction. Even in two-dimensional simulations, it was not
possible to observe electrons beyond 200 ${\mev}$, as measured in this experiment, since except in three-dimensional simulations, both the
radial plasma wave enhancement and self-focusing effects are underestimated. Hence it is only in three-dimensional simulations that
$E_\mathrm{max} \sim E_\mathrm{wb}$ can be reached. That such large electric fields are generated demonstrates another important difference between FLW and SMWF regimes, since in the latter, plasma heating by instabilities limits the accelerating electric field to
an order of magnitude below the cold wave-breaking limit. It should be noted that the peak electric field inferred for these FLW
experiments is in excess of 1 ${\tvm}$, considerably larger than any other coherent accelerating structure created in the
laboratory.

   \begin{figure}
    \begin{center}
     \includegraphics[width=8cm]{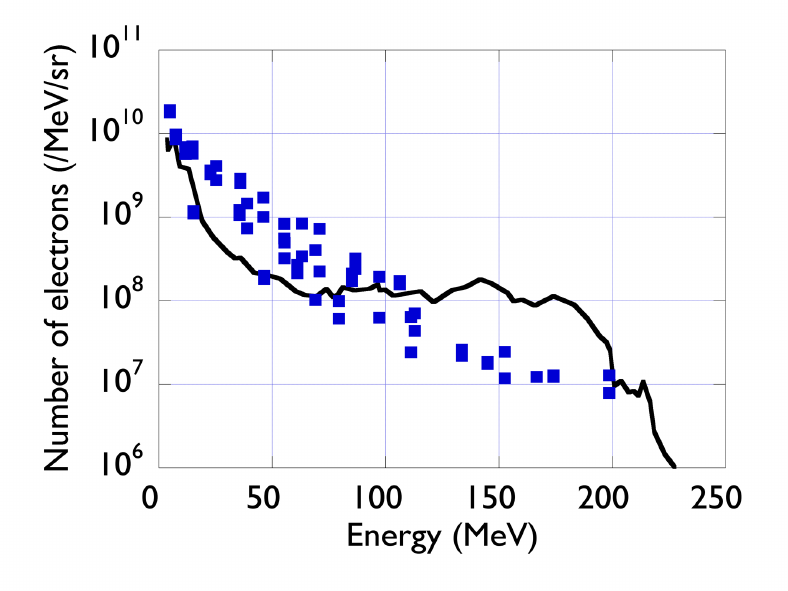}
    \end{center}
    \caption{Typical experimental (blue squares) and calculated (black curve) electron spectrum obtained at $n_\mathrm{e}=~2.5~\times 10^{19}$ cm$^{-3}$ with a 1 J, 30 fs laser pulse focused down to a waist of ${w}_0= 18 \mu$m.}
\label{FigSpectreMaxwellien}
   \end{figure}

  \subsubsection{Bubble regime} \label{SecRegimeBulle}
\begin{figure}[htbp]
    \begin{center}
     \includegraphics[width=12cm]{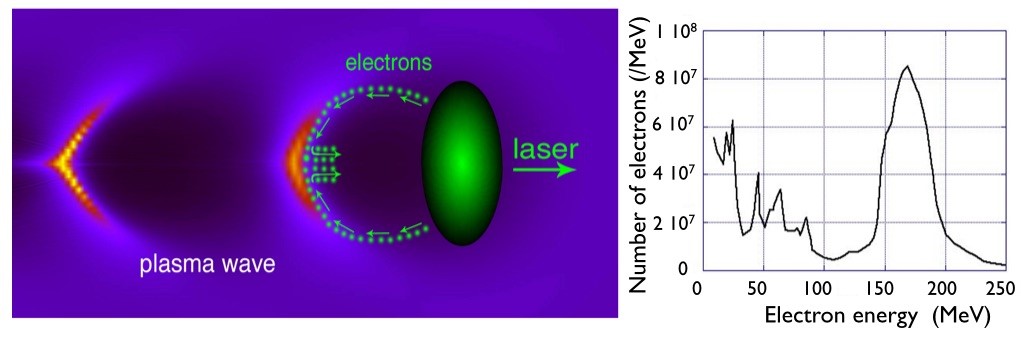}
    \end{center}
    \caption{Left: acceleration principle in the bubble regime. Right: typical quasi-monoenergetic electron spectra measured at LOA.}
   \label{FigRegimeBulle}
   \end{figure}

   In 2002, theoretical work based on three-dimensional particle-in-cell (PIC) simulations have shown the existence of a robust acceleration mechanism called the bubble regime \cite{pukh02}. In this regime, the dimensions of the focused laser are shorter than the plasma wavelength in longitudinal and also transverse directions, the laser shape appearing like a ball of light. If the laser energy contained in this spherical volume is large enough, the ponderomotive force of the laser expels radially and efficiently electrons from the plasma, which forms a cavity free from electrons behind the laser surrounded by a dense region of electrons. Behind the bubble, electron trajectories intersect each other. Electrons are injected into the cavity and accelerated along the laser axis, thus creating an electron beam with radial and longitudinal dimensions smaller than those of the laser (see Fig. \ref{FigRegimeBulle}).

   The signature of this regime is a quasi-monoenergetic electron distribution that results from the localization of injection at the back of the cavity, which gives similar initial properties in the phase space to injected electrons. Since electrons are trapped behind the laser pulse, this reduces or even suppresses interaction with the electric field of the laser. The trapping process stops when the charge contained in the cavity compensates the ionic charge, and the rotation in the phase space also leads to a shortening of the spectral width of the electron beam \cite{tsun04}.

   Several laboratories have obtained quasi-monoenergetic spectra: in France \cite{faur04} with a laser pulse shorter than the plasma period, but also with pulses slightly longer than the plasma period in the U.K. \cite{mang04}, in the United States \cite{gedd04}, then in Japan \cite{miur05} and in Germany \cite{hidd06}. The interest in such a beam is a result of its importance for a number of applications: it is now possible to transport and to refocus this beam by magnetic fields. With a Maxwellian-like spectrum, it would have been necessary to select an energy range for the transport, which would have decreased significantly the electron flux. Electrons in the ${\gev}$ level were also observed in this regime using a uniform plasma \cite{hafz08} or in plasma discharge, i.e. a plasma with a parabolic density profile \cite{leem06} with a more powerful laser which propagates at high intensity over a longer distance. With the development of PW class lasers, a few ${\gev}$ electron beam has been reported \cite{leem14, hyun13, wang13}.

\begin{figure}[htbp]
   \begin{center}
\includegraphics[width=14cm]{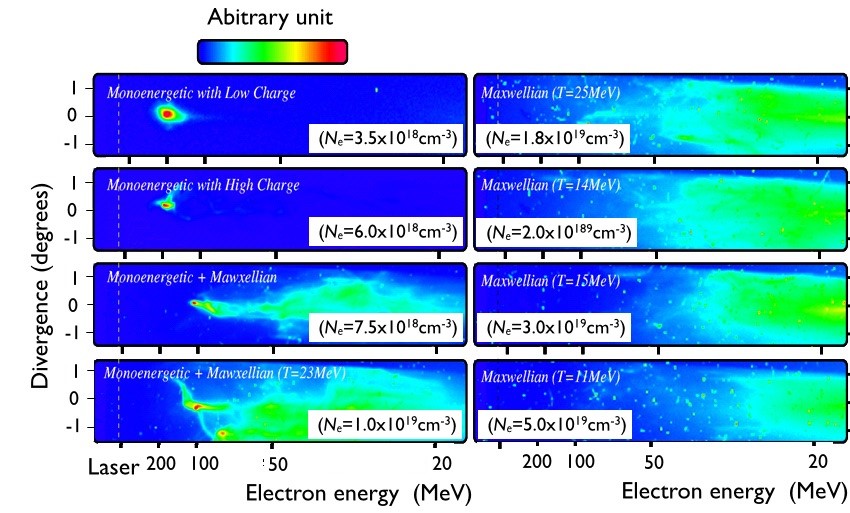}
\caption{\label{FigTransition} Electron beam distribution for different plasma densities showing the transition from the self-modulated laser wakefield and the forced laser wakefield to the bubble/blow-out regime. From top to bottom, the plasma density values are $6 \times10^{18}\cc$, $1 \times10^{19}\cc$, $2 \times10^{19}\cc$ and $5 \times10^{19}\cc$.}
\end{center}
\end{figure}

In all the experiments performed so far, the laser plasma parameters were not sufficient to fully enter the bubble/blowout regimes. Yet, with the increase of laser system power, this regime will be reached, and significant improvement of the reproducibility of the electron beam is expected. Nevertheless, since self-injection occurs through transverse wave-breaking, it is hardly appropriate for a fine tuning and control of the injected electron bunch.
Figure \ref{FigTransition} shows electron distributions obtained for different densities. It illustrates the transition from a Maxwellian-like spectrum obtained in high density cases in the self-modulated laser wakefield, to the forced laser wakefield regime with an emerging monoenergetic component at moderate density, to a spectrum containing a very well defined monoenergetic component. This transition occurs for densities around $n_\mathrm{e}= $1--3 $\times 10^{19}$ cm$^{-3}$. The best coupling for obtaining a high charge and a quasi-monoenergetic electron beam is at $n_\mathrm{e}=6\times 10^{18}$ cm$^{-3}$. For this density, the image shows a narrow peak around 170 ${\mev}$, indicating efficient monoenergetic acceleration with a 24\% energy spread corresponding
to the spectrometer resolution.

 \subsection{Injection in a density gradient}
One solution to control electron injection with current laser technology was proposed by S. Bulanov~$et$~$al.$ ~\cite{bula98}. It involves a downward density ramp with a density gradient scale length $L_\mathrm{grad}$ smaller than the plasma wavelength $\lambda_\mathrm{p}$.
Injection in a downward density ramp relies on the slowing down of the plasma wave velocity at the density ramp. This decrease of the plasma wave phase velocity lowers the threshold for trapping the plasma background electrons and causes wave-breaking of the wakefield in the density ramp. This method can therefore trigger wave-breaking in a localized spatial region of the plasma. Geddes \emph{et al.} ~\cite{gedd08} showed the injection and acceleration of high charge (${>}300$ pC) and stable quality beams of ${\simeq}0.4$${\mev}$ in the downward density ramp at the exit of a gas jet ($L_\mathrm{grad}\simeq 100\mic \gg \lambda_\mathrm{p}$). These results, although very promising, have the disadvantage that the low energy beam blows up very quickly out of the plasma, due to the space-charge effect. To circumvent this issue, one should use a density gradient located early enough along the laser pulse propagation so that electrons can be accelerated to relativistic energies ~\cite{kim04}. This can be achieved by using, for instance, a secondary laser pulse to generate a plasma channel transverse to the main pulse propagation axis~\cite{chie05}. In this case, the electron beam energy could be tuned by changing the position of the density gradient. In this pioneering experiment, the electron beam had a large divergence and a Maxwellian energy distribution because of a too low laser energy. However, two-dimensional PIC simulations showed that this method can result in high quality quasi-monoenergetic electron beams ~\cite{toma03}.

	At LOA a density gradient across a laser created plasma channel was used to stabilize the injection ~\cite{faur10}. The experiment was performed at an electron density close to the resonant density for the laser wakefield ($c\tau\sim\lambda_\mathrm{p}$) to guaranty a post acceleration that delivered high quality electron beams with narrow divergences (4 mrad) and quasi-monoenergetic electron distributions with 50 to 100 pC charge and 10\% relative energy spread.

The use of density gradients at the edges of a plasma channel showed an improvement of the beam quality and of the reproducibility with respect to those produced in the bubble/blowout regime with the same laser system and with similar laser parameters. However, the electron energy distribution was still found to fluctuate from shot to shot.
The performance of the experiments could be further improved and could potentially lead to more stable and controllable high quality electron beams. In particular, sharper gradients with $L_\mathrm{grad}\simeq\lambda_\mathrm{p}$ coupled with a long plasma can lead to better beam quality~\cite{bran08}.

 \begin{figure}[t]
\begin{center}
\includegraphics[width=16cm]{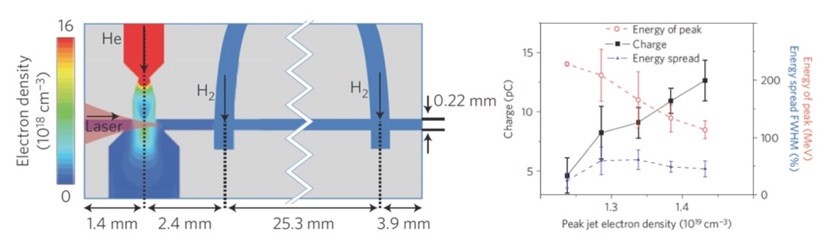}
\caption{\label{fjetcap} Top: target schematic representation with embedded supersonic gas jet into a capillary that is filled with hydrogen gas. Bottom: the charge (squares), energy (circles) and energy spread (triangles) as a function of the peak jet density. From A. J. Gonsalves \textit{et al.}~\cite{gons11}.}
\end{center}
\end{figure}

For example, at LBNL, as shown in Fig.~\ref{fjetcap}, electrons at 30 ${\mev}$ were produced in a density ramp and accelerated up to 400 ${\mev}$ in a second stage 4 cm parabolic plasma channel formed with a plasma discharge~\cite{gons11}. Here also, the density gradient injection led to an improvement of the stability and of the electron beam quality. The electron energy, divergence, charge and relative energy spread were found to be respectively 400 ${\mev}$, 2 mrad, 10 pC and 11\%.
It was shown that steeper density transitions, with $L_\mathrm{grad} \ll \lambda_\mathrm{p}$, can also cause trapping~\cite{suk01}. Such injection was successfully demonstrated experimentally using the shock-front created by a knife-edge inserted in a gas jet ~\cite{Koya09,Schm10}.

\begin{figure}[t]
  \begin{center}
\includegraphics[width=14cm]{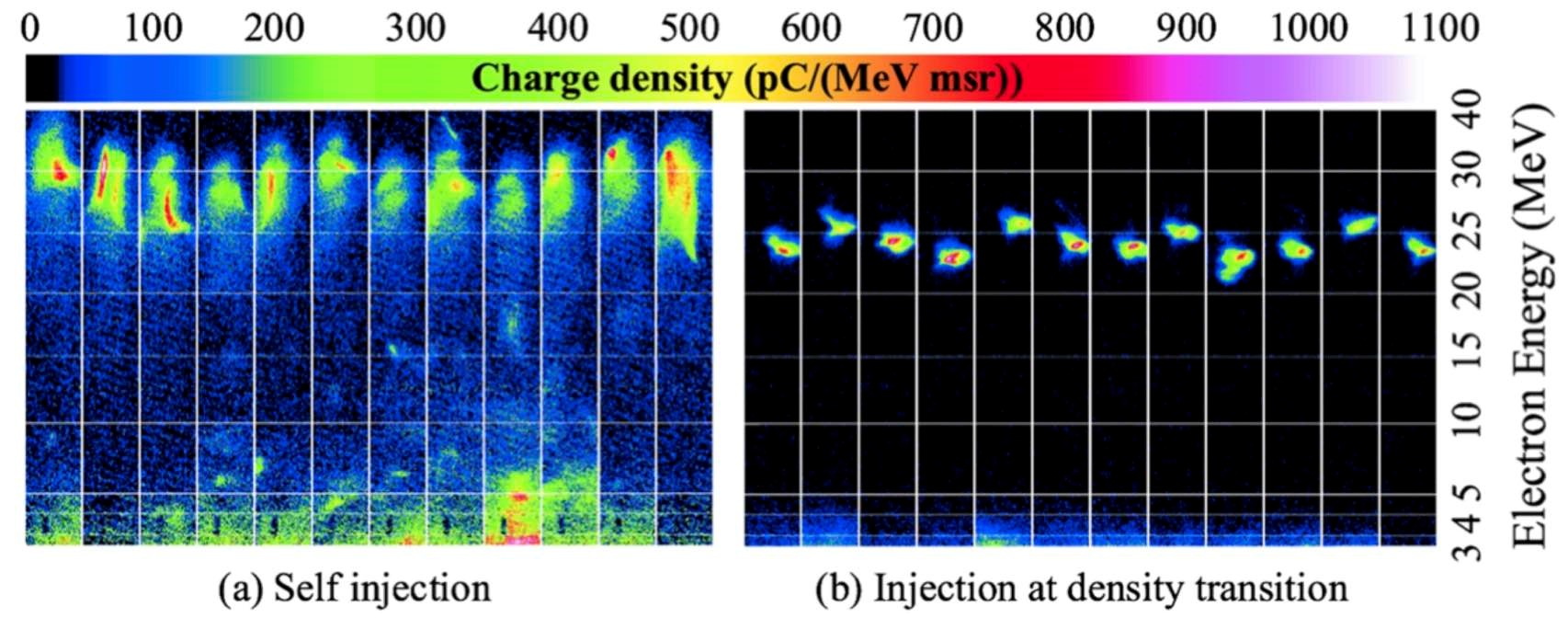}
\caption{\label{fig2gradient} A few representative shots of the 10\% of all the shots with the lowest energy spread for self-injection (top) and injection at a density transition (bottom). The horizontal axis in each image corresponds to the transverse electron beam size; the vertical axis shows electron energy. From K. Schmid \textit{et al.}\cite{Schm10}.}
\end{center}
\end{figure}

 Figure \ref{fig2gradient} illustrates the improvement of injection in a sharp density gradient, with a characteristic length of the order of the plasma wavelength and a peak electron density of about $5 \times10^{19} \cc$. The experiment was performed at the Max-Planck-Institut fur Quantenoptik using a multi-TW sub-10 fs laser system that delivered for this experiment pulses with 65 mJ energy on target and with a duration of 8~fs~FWHM. The laser pulse was focused down to a spot diameter of $12\mic$ FWHM into the gas target yielding a peak intensity of $2.5\times10^{18} \wcm$. The comparison between the self-injection and density transition injection shows a reduction of the relative energy spread and of the charge of a about a factor of 2.

 \subsection{Injection with colliding laser pulses}

\begin{figure}[t]
  \begin{center}
\includegraphics[width=14.8cm]{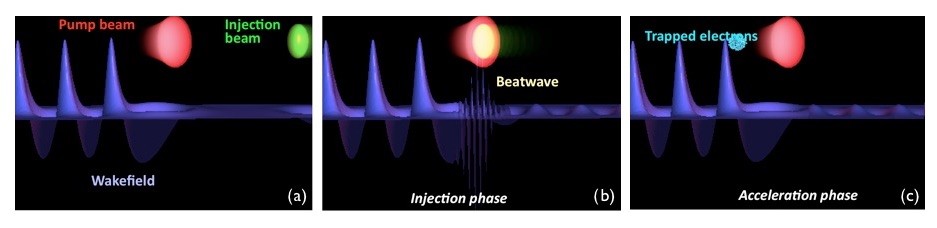}
\caption{\label{fig3colliding_principle} Scheme of the principle of the injection with colliding laser pulses: (a) the two laser pulses propagate in opposite direction; (b) during the collision, some electrons get enough longitudinal momentum to be trapped by the relativistic plasma wave driven by the pump beam and (c) trapped electrons are then accelerated in the wake of the pump laser pulse.}
\end{center}
\end{figure}

 In 2006, stable and tunable quasi-monoenergetic electron beams were measured by using two counterpropagating laser beams in the colliding scheme. The use of two laser beams instead of one offers more flexibility and enables one to separate the injection from the acceleration process~\cite{faur06}. The first laser pulse, the pump pulse, is used to excite the wakefield while the second pulse, the injection pulse, is used to heat electrons during the collision with the pump pulse. After the collision has occurred, electrons are trapped and further accelerated in the wakefield, as shown in Fig.~\ref{fig3colliding_principle}.

To trap electrons in a regime where self-trapping does not occur, one has either to inject electrons with energies greater that the trapping energy or dephase electrons with respect to the plasma wave. As mentioned earlier, electrons need to be injected in a very short time (${\ll}\lambda_\mathrm{p}/c$) in order to produce a monoenergetic beam. This can be achieved using additional ultra-short laser pulses whose only purpose is to trigger electron injection.

Umstadter $et$ $al.$ ~\cite{umst96} first proposed to use a second laser pulse propagating perpendicular to the pump laser pulse. The idea was to use the radial ponderomotive kick of the second pulse to inject electrons. Esarey $et$ $al.$~\cite{esar97} proposed a counter-propagating geometry based on the use of three laser pulses. This idea was further developed by considering the use of two laser pulses~ \cite{fubi04}. In this scheme, a main pulse (pump pulse) creates a high amplitude plasma wave and collides with a secondary pulse of lower intensity. The interference of the two beams creates a beatwave pattern, with a zero phase velocity, that heats some electrons from the plasma background. The force associated with this ponderomotive beatwave is proportional to the laser frequency. It is therefore many times greater than the ponderomotive force associated with the pump laser, that is inversely proportional to the pulse duration at resonance. As a result, the mechanism is still efficient even for modest laser intensities. Upon interacting with this field pattern, some background electrons gain enough momentum to be trapped in the main plasma wave and then accelerated to high energies. As the overlapping of the lasers is short in time, the electrons are injected over a very short distance and can be accelerated to an almost monoenergetic beam.

This concept was validated in an experiment ~\cite{faur06}, using two counter-propagating pulses. Each pulse had a duration of 30 fs FWHM, with $a_0=1.3$ and $a_1=0.4$. They were propagated in a plasma with electron density $n_\mathrm{e}=7 \times 10^{18}$ cm$^{-3}$ corresponding to $\gamma_\mathrm{p}=k_0/k_\mathrm{p}=15$. It was shown that the collision of the two lasers could lead to the generation of stable quasi-monoenergetic electron beams. The beam energy could be tuned by changing the collision position in the plasma.

PIC simulations in one dimension have been used to model electron injection in the plasma wave at the collision of the two lasers,
and their subsequent acceleration. In particular, the PIC simulations were compared to existing fluid models~\cite{esar97} with prescribed electric field. They showed significant differences, such as changes in the behaviour of plasma fields and in the amount of injected charge. The fluid approach fails to describe qualitatively and quantitatively many of the physical mechanisms that occur during and after the laser beams collision ~\cite{rech07}. In this approach, the electron beam charge was found to be one order of magnitude greater than in the PIC simulations. For a correct description of injection, one has to describe properly (i) the heating process, e.g. kinetic effects and their consequences on the dynamics of the plasma wave during the beating of the two laser pulses and (ii) the laser pulse evolution which governs the dynamics of the relativistic plasma waves ~\cite{davo08}. Unexpectedly, it was shown that efficient stochastic heating can be achieved when the two laser pulses are crossed polarized. The stochastic heating can be explained by the fact that for high laser intensities, the electron motion becomes relativistic which introduces a longitudinal component through the $\mathbf{v}\times \mathbf{B}$ force. This relativistic coupling makes it possible to heat electrons even in the case of crossed polarized laser pulses ~\cite{malk09}. Thus the two perpendicular laser fields couple through the relativistic longitudinal motion of electrons. The heating level is modified by tuning the intensity of the injection laser beam or by changing the relative polarization of the two laser pulses ~\cite{rech09c}. This consequently changes the volume in the phase space of the injected electrons and therefore the charge and the energy spread of the electron beam.

\begin{figure}[t]
\begin{center}
\includegraphics[width=8cm]{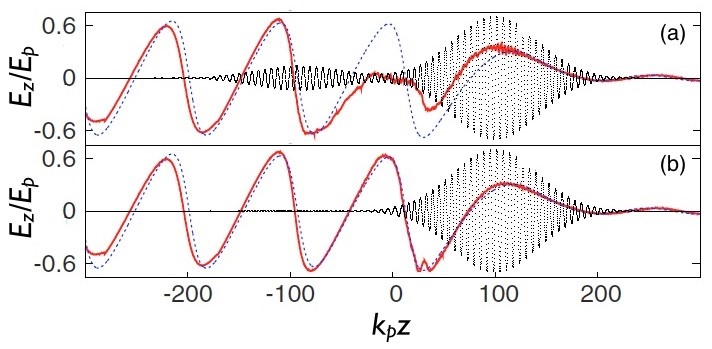}
\caption{\label{diffcollipar_perp} Longitudinal electric field computed at $t=43$ fs in one-dimensional PIC simulation (solid red line), and in fluid simulations (dotted blue line). The transverse electric field is also represented (thin dotted line). The laser pulse duration is 30 fs FWHM, the wavelength is 0.8 $\mu$m with $a_0=2$ and $a_1=0.4$. The laser pulses propagate in a plasma with electron density $n_\mathrm{e}=7 \times 10^{18}$ cm$^{-3}$. In (a) the case of parallel polarization and in (b) the case of crossed polarization.}
\end{center}
\end{figure}

Figure \ref{diffcollipar_perp} shows, at a given time (42 fs), the longitudinal electric field during and after collision for parallel and crossed polarization. The solid line corresponds to PIC simulation results whereas the dotted line corresponds to fluid calculations. The laser fields are represented by the thin dotted line. When the pulses have the same polarization, electrons are trapped spatially in the beatwave and cannot sustain the collective plasma oscillation, inducing a strong inhibition of the plasma wave which persists after the collision. When the polarizations are crossed, the electron motion is only slightly disturbed compared to their fluid motion, and the plasma wave is almost unaffected during the collision, which tends to facilitate trapping.

\begin{figure}[t]
\begin{center}
\includegraphics[width=8cm]{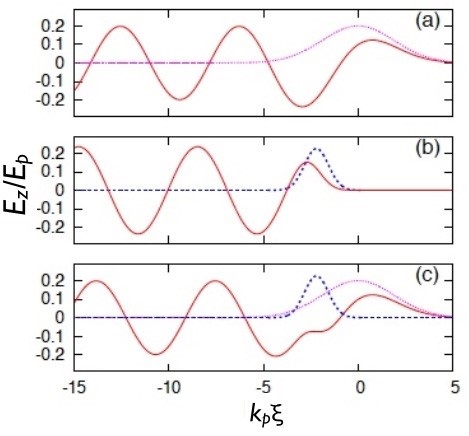}
\caption{\label{beamload} In red, normalized longitudinal electric field. (a) The laser (in pink) wakefield. (b) The electron bunch (in blue) wakefield. (c) Field resulting from the superposition of the laser and electron beam wakefields. The normalized vector potential is $a_0=1$, the laser pulse duration is 30 fs, $n_\mathrm{e}=7 \times 10^{18}$ cm$^{-3}$, $n_\mathrm{beam}=0.11\times n_\mathrm{e}$, the bunch duration is 10 fs and its diameter is 4 $\mu$m. From C. Rechatin, Ph.D. thesis.}
\end{center}
\end{figure}

Importantly, it was shown that the colliding pulse approach allows control of the electron beam energy which is done simply by changing the delay between the two laser pulses ~\cite{faur06}. The robustness of this scheme permitted also very accurate studies of the dynamics of the electric field in presence of a high current electron beam to be carried out.
Indeed, in addition to the wakefield produced by the laser pulse, a high current electron beam can also drive its own wakefield as shown in Fig.~\ref{beamload}.

The beam loading effect contributes to the reduction of the relative energy spread of the electron beam. It was demonstrated that there is an optimal load which flattened the electric field, leading to the acceleration of all the electrons with the same value of the field, and producing consequently an electron beam with a very small, 1\%, relative energy spread ~\cite{rech09b}. Thanks to the beam loading effect, the most energetic electrons can be slightly slowed down and accelerated to the same energy as the slowest ones. In cases of low charge beam, this effect does not play any role and the energy spread depends mainly on the heated volume. For a very high current, the load is too high and the most energetic electrons slow down so much that they eventually obtain energies smaller than the slowest electrons~\cite{rech09b}, increasing the relative energy spread. The existence of an optimal load was observed experimentally and supported by full three-dimensional PIC simulations. It corresponds to a peak current in the 20--40 kA range. The decelerating electric field due to the electron beam was found to be in the ${\mathrm{GV/m/pC}}$ range.

\subsection{Injection triggered by ionization}
Another scheme was proposed recently to control the injection by using a high-$Z$ gas and/or a high-$Z$/low-$Z$ gas mixture. Thanks to the large differences in ionization potentials between successive ionization states of the atoms, a single laser pulse can ionize the low energy level electrons in its leading edge, drive relativistic plasma waves, and inject in the wakefield the inner level electrons which are ionized when the laser intensity is close to its maximum.

\begin{figure}[t]
\begin{center}
\includegraphics[width=14cm]{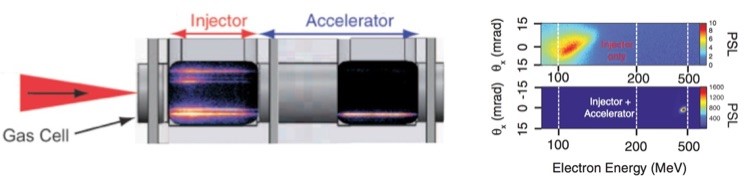}
\caption{\label{twostagel} (a) Schematic of the experimental set-up showing the laser beam, the two-stages gas cell, on the left the injector part and on the right the accelerator part. (b) Magnetically dispersed electron beam images from a 4 mm injector-only gas cell (top) and the 8 mm two-stages cell (bottom). From B. B. Pollock \textit{et al.} ~\cite{Clay11}.}
\end{center}
\end{figure}

Such an ionization trapping mechanism was first demonstrated in electron beam driven plasma wave experiments on the Stanford Linear Collider (SLAC) ~\cite{oz07}. Electron trapping from ionization of high-$Z$ ions from capillary walls was also inferred in experiments on laser wakefield acceleration ~\cite{Rowl08}. In the case of a self-guided laser driven wakefield, a mixture of helium and trace amounts of different gases was used~\cite{Pak10,Guff10}. In one of these experiments, electrons from the K shell of nitrogen were tunnel ionized near the peak of the laser pulse and were injected into and trapped by the wake created by electrons from majority helium atoms and the L shell of nitrogen. Because of the relativistic self-focusing effect, the laser propagates over a long distance with peak intensity variations that can trigger the injection over a long distance and in an inhomogeneous way, which leads to the production of a high relative energy spread electron beam. Importantly, the energy required to trap electrons is reduced, making this approach of great interest to produce electron beams with a large charge at moderate laser energy. To reduce the distance over which electrons are injected, experiments using two gas cells were performed at LLNL ~\cite{Clay11}, as shown in Fig.~\ref{twostagel}. By restricting electron injection to a small region, in a first short cell filled with a gas mixture (the injector stage), energetic electron beams (of the order of 100 ${\mev}$) with a relatively large energy spread were generated. Some of these electrons were then further accelerated in a second, larger, accelerator stage, consisting of a long cell filled with low-Z gas, which increases their energy up to 0.5 ${\gev}$ while reducing the relative energy spread to ${<}5\% $ FWHM.

\subsection{Longitudinal injection}
\begin{figure}[t]
\begin{center}
\includegraphics[width=14cm]{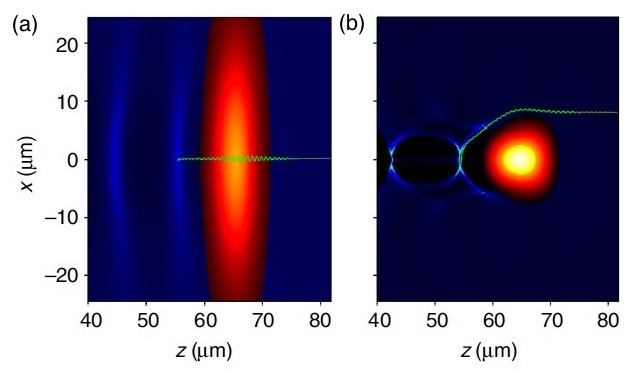}
\caption{\label{long} Schematic for longitudinal and transverse self-injections. (a) Typical trajectory of an injected electron in the longitudinal self-injection
mechanism. (b) Typical trajectory of an injected electron in the transverse self-injection mechanism. The blue colour scale represents the electron
density. The red to yellow colour scale indicates the laser intensity. The trajectories are given by the green lines.}
\end{center}
\end{figure}

As has been shown, electron trapping is generally achieved by the wave-breaking of the plasma wake, a process that is by nature
uncontrollable and leads generally to poor quality electrons. The presented controlled injection techniques, such as colliding pulse injection, ionization-induced injection and density gradient injection, have been developed to overcome this shortcoming.
These methods offer an improved control on the acceleration and lead to better electron features, but they imply generally
complex set-ups. For this reason, self-injection remains the most common method for injecting electrons in the plasma wake. Two
distinct physical mechanisms can be distinguished: longitudinal and transverse self-injection. In longitudinal self-injection, the
trajectory of injected electrons is mainly longitudinal, with a negligible transverse motion. As shown by the schematic in
Fig.~\ref{long}, the injected electrons pass through the laser pulse and gain energy while crossing the plasma wave. When they reach the
rear of the first plasma period, their velocity exceeds the wake phase velocity and the electrons are eventually injected. The only
electrons that are trapped are those that were initially close to the axis where the laser intensity and the wakefield
amplitude are the highest and where the ponderomotive force is small. The longitudinal self-injection mechanism is analogous
to one-dimensional longitudinal wave-breaking. In contrast, transverse self-injection occurs in the bubble regime, where the laser ponderomotive force expels electrons from the propagation axis and forms an electron-free cavity in its wake. As shown in Fig.~\ref{long} and in Fig.~\ref{FigRegimeBulle}, the injected electrons are initially located at approximately one laser waist from the axis. They circulate around the laser pulse and the bubble, and attain a velocity larger than the wake phase velocity when reaching the axis at the rear of the bubble.

During its propagation, the laser pulse evolves, the self-phase-modulation modifies its duration and the relativistic self-focusing modifies its initial transverse shape. As a consequence, the generated wakefield is not uniform along the laser propagation axis and electrons can be self-injected at different positions of the plasma accelerator. Electrons in the second bunch originate from positions close to the laser waist, as
expected in the case of trapping by transverse self-injection. In contrast, electrons in the first bunch come from regions close to
the axis. When these electrons are injected, the laser spot radius is large and the normalized laser amplitude is still low; hence, the radial ponderomotive force close to the axis is small.

\begin{figure}[t]
\begin{center}
\includegraphics[width=14cm]{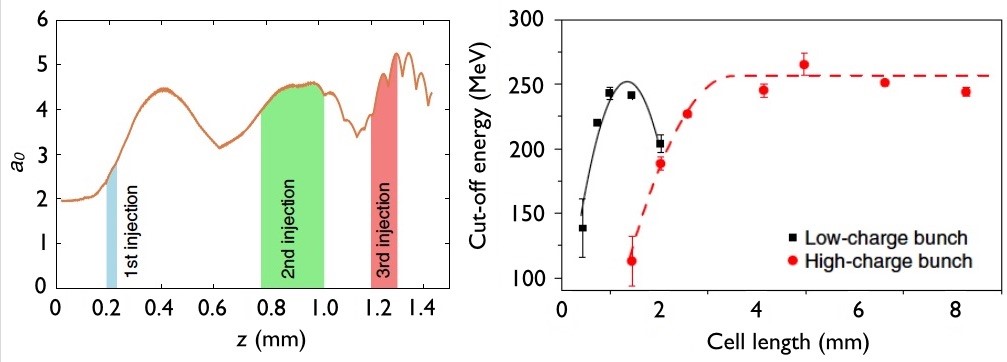}
\caption{\label{logi} Left: Evolution of the normalized laser amplitude. The coloured areas indicate the injection regions. Right: Electron spectra for a density of $1.1 \times10^{19}\cc$. Mean $5\%$ cut-off energy as a function of the cell length. The solid line is a parabolic fit, the dashed line is a visual guide.
The peak accelerating electric field is $ E_\mathrm{max}^{z}=340 \pm 65 {\gvm}$ for the first bunch and $E_\mathrm{max}^{z}=185 {\pm} 40 {\gvm}$ for the second bunch.}
\end{center}
\end{figure}

Thus, on-axis electrons are only weakly deviated when crossing the laser pulse, and they remain in
the region of largest accelerating field $E_{z}$. Moreover, the laser amplitude increases steeply in the region of first injection because
of laser self-focusing (see Fig.~\ref{logi}). This reduces the wake phase velocity via the relativistic shift of the plasma wavelength $\lambda_\mathrm{p}$ ~\cite{Kalm10}. The strongly reduced wake phase velocity lowers the threshold for trapping, such that electrons can catch up the plasma wave
and be injected despite a low $a_{0}$, similarly to density gradient injection.

  \section{Particle beam wakefield accceleration}

 \subsection{Electron driven plasma wakefield}

\begin{figure}[t]
\begin{center}
\includegraphics[width=14cm]{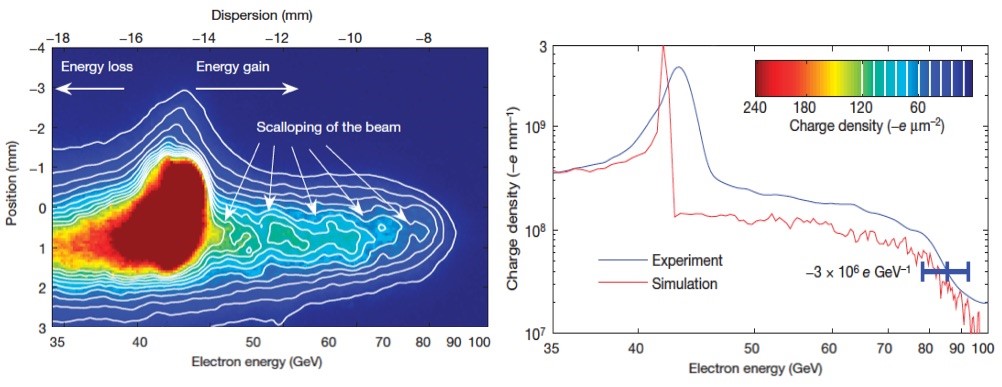}
\caption{\label{nature07} Energy spectrum of the electrons in the 35--100 ${\gev}$ range. The dispersion
(shown on the top axis) is inversely proportional to the particle energy (shown on the bottom axis). The head of the pulse, which is unaffected by the
plasma, is at 43 ${\gev}$. The core of the pulse, which has lost energy driving the plasma wake, is dispersed partly out of the field of view of the camera.
Particles in the back of the bunch, which have reached energies up to 85 ${\gev}$, are visible to the right.}
\end{center}
\end{figure}

Wakefields in a plasma can be also driven by an electron bunch that has, at resonance, a length of about half the plasma wavelength. Whereas in the laser wakefield case the radiation pressure, known as the ponderomotive force, pushes away the plasma electrons, here the force is due the space-charge of the electron beam. The plasma electrons are strongly blown out radially, but because of the space-charge attraction of the plasma ions, they are attracted back towards the rear of the beam where they overshoot the beam axis and set up a wakefield oscillation. Here again, charged particles injected in an appropriately phased trailing pulse can then extract energy from the wakefield. Because of the lack of accelerators that deliver suitable electron beams, there are fewer particle-beam-driven plasma acceleration experiments compared with laser accelerator experiments. The first beam-driven plasma wakefield experiments were carried out at the Argonne Wakefield Accelerator Facility in the 1980s ~\cite{rose88a}. Since then, important experiments done at SLAC by the UCLA/USC/SLAC collaboration have mapped the physics of electron and positron beam-driven wakes and has shown acceleration gradients of $40 {\gvm}$ using electron beams with metre-scale plasmas ~\cite{blum07}.
In the first important SLAC experiments only one electron bunch was used to excite the wakefield. Since the energy of the drive pulse was 42 ${\gev}$, both the electrons and the wake are moving at a velocity close to c, so there is no relative motion between the electrons and the wakefield. Because the electron bunch was also longer that the plasma period, most of the electrons in the drive bunch lose energy in exciting the wake, but some electrons in the back have gained energy from the wakefield as the wakefield changes its sign. Thanks to the high quality, low emittance of the electron bunch, its intensity was so high that the 42 ${\gev}$ electron beam passed through a column of lithium vapour 85 cm long, the head of the beam created a fully ionized plasma and the remainder of the beam excited a strong wakefield. Figure ~\ref{nature07} shows the energy spectrum of the beam measured after the plasma. The electrons in the bulk of the pulse that lost energy in driving the wake are mostly dispersed out of the field of view of the spectrometer camera and so are not seen in the spectrum. However, electrons in the back of the same pulse are accelerated and reach energies up to 85 ${\gev}$. The measured spectrum of the accelerated particles was in good agreement with the spectrum obtained from computer simulations of the experiment, as Fig. ~\ref{nature07} shows. As said Prof. C. Joshi, `This is a remarkable result when one realizes that while it takes the full 3 km length of the SLAC linac to accelerate electrons to 42 ${\gev}$, some of these electrons can be made to double their energy in less than a metre'.

\begin{figure}[t]
\begin{center}
\includegraphics[width=12cm]{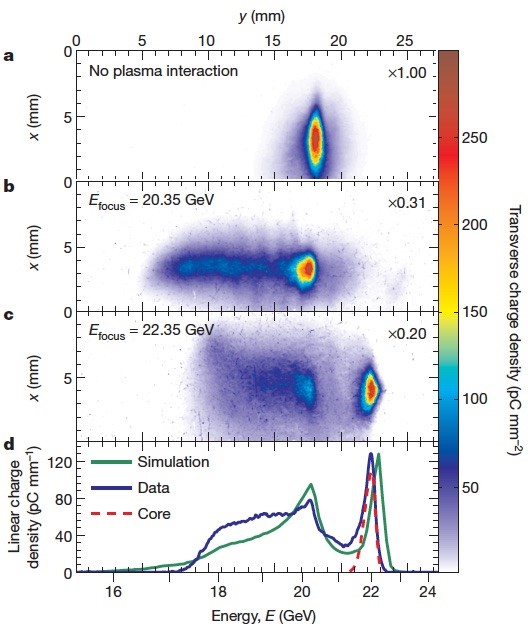}
\caption{\label{nature14} (a) The dispersed electron beam profile without plasma interaction, where the spectrometer is set to image
22.35 ${\gev}$. (b) and (c) The dispersed beam profile after the electron bunches have interacted with the plasma, where the spectrometer is set to image 20.35 ${\gev}$ and 22.35 ${\gev}$. (d) The spatially integrated spectrum (in $x$) or the linear charge density of the bunches shown in (c) (solid blue line) along with
the final spectrum obtained from the simulation (solid green line in (d)). The core of the accelerated trailing beam is shown for the data
(dashed red line).}
\end{center}
\end{figure}

In this former experiment, a small fraction of electrons of the beam was injected and accelerated. As a consequence, the quality of the accelerated electrons was poor with a long Maxwellian-like tail and therefore with also a poor energy transfer efficiency. For future high energy physics purposes, a high efficiency is mandatory to achieve an affordable and compact high-energy collider. To improve this, a second important experiment has been performed at SLAC. In this plasma wakefield acceleration experiment, a charge-density wake with high accelerating fields has been driven by an ultra-relativistic bunch of charged particles (the drive bunch) through a plasma followed by a second bunch of relativistic electrons (the trailing bunch) in the wake of the drive bunch at an appropriate distance that has been efficiently accelerated to higher energy. Whereas in the previous experiment, the total charge of accelerated electrons was insufficient to extract a substantial amount of energy from the wake, here high efficiency acceleration of the trailing bunch of electrons has been demonstrated. Accelerations of approximately 70--80 pC of the trailing bunch have been achieved in an accelerating gradient of about 4.4 ${\gvm}$. As presented in Fig.~\ref{nature14}, these particles have gained approximately 1.6 ${\gev}$ of energy per particle, with a final energy spread as low as $1\%$ and an energy-transfer efficiency from the wake to the bunch that exceeded $30\%$. This acceleration of a distinct bunch of electrons containing a substantial charge and having a small energy spread with both a high accelerating gradient and a high energy-transfer efficiency represents a milestone in the development of plasma wakefield acceleration into a compact and affordable accelerator technology. $6\%$ of the initial electron beam energy (36 J) was transferred to the trailing bunch. This value is comparable to the laser to electron beam energy transfer efficiency from LPAW. The main advantage here being that the driver is more efficient that the laser driver. Accelerators have indeed today a wall-plug efficiency more that 10 times larger than lasers.

  \subsection{Proton driven plasma wakefield}

As it has been shown in all these former experiments, the energy gain was limited by the energy carried by the driver (about 40 J for an e-beam driver and about 100 J for a laser driver) and by the propagation length of the driver in the plasma (few tens of centimetres for the e-driver and few centimetres for the laser driver). The laser pulse and electron bunch driver schemes therefore require the use of many acceleration stages in the tens of ${\gev}$ each in order to gain ${\tev}$ energy levels. A 10 ${\gev}$ stage that delivers an nC of charge corresponds to an energy of 10 J, and it will correspond to 10 kJ for a 10 ${\tev}$ stage. If one assumes $10\%$ energy transfer efficiency from the driver to the trail bunch, this indicates that the driver energy must contain about 100 kJ. In 2009, for the first time, plasma-wake excitation by a relativistic proton beam has been considered~\cite{cald09}. In this ideal case, the proton driver beam has to be resonant with the plasma and it was predicted on the basis of numerical simulations that 10 ${\gev}$ electrons injected could be accelerated to 0.5 ${\tev}$ in a 450 m proton wakefield. Unfortunately such a short proton bunch does not exist, and therefore it has been proposed to use the CERN SPS 19 kJ, 400 ${\gev}$ proton beam that is produced routinely.

\begin{figure}[t]
\begin{center}
\includegraphics[width=14cm]{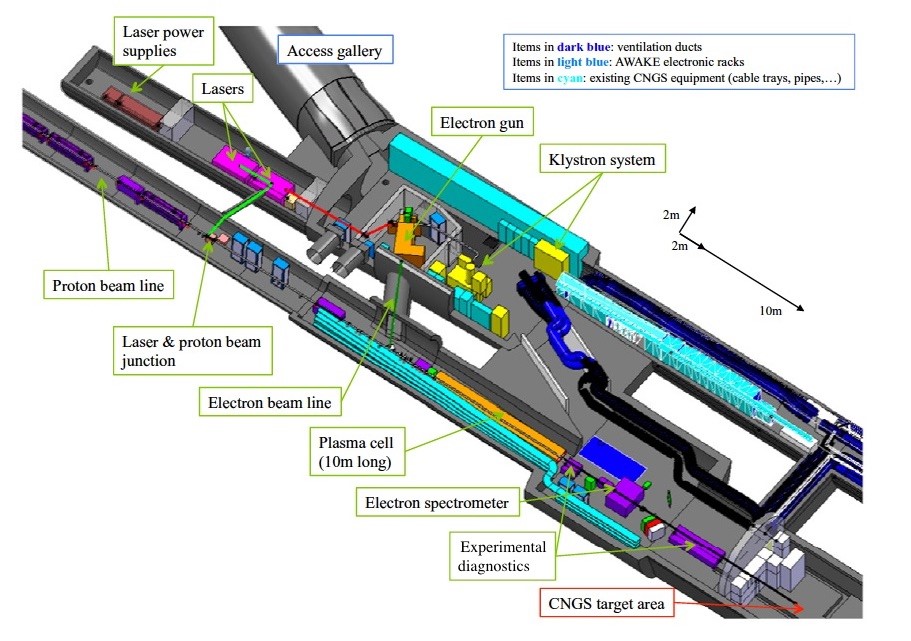}
\caption{\label{awa} Design of the layout of the AWAKE experiment}
\end{center}
\end{figure}

Because the length of the driver (about 10 cm) is much longer than the plasma wavelength (about 1~mm) at a density large enough (of $10^{14}$ cm$^{-3}$) to reach a ${\gvm}$ accelerating field, the interaction has to occur in the self-modulated regime. In this regime, also called the self-modulation instability (SMI)~\cite{kuma10}, the proton bunch is split during its propagation in several micro-bunches that excite resonantly a strong plasma wave. A first experiment, called AWAKE ~\cite{assm14}, will be performed at CERN in the next two years to demonstrate the possibility of accelerating injected electrons in proton driven plasma wakefield. For this, an extremely uniform long plasma has been developed with a control precision for the density that must be within $0.5\%$ over metres. To optimize the coupling efficiency, electrons will be injected at an angle after the self-modulated instability has reached saturation. The conceptual design of the proposed AWAKE experiment is shown in Fig.~\ref{awa} where the laser and proton bunches are made co-linear. The laser that will ionize the metal vapour is required to seed the self-modulation instability. The self-modulated proton bunch (shown on the left hand side) enters a second plasma section where it drives the plasma wakefield structure (shown on the right side). The electrons are injected in the wakefields and the 2~${\gev}$ accelerated electrons will be measured with an electron spectrometer. The AWAKE experiment will be installed in the CERN Neutrinos to Gran Sasso (CNGS) facility. Approximately $5\%$ of electrons are supposed to be trapped and accelerated to the end of the 10 m plasma with accelerating gradients in the few ${\gvm}$. In addition to the electron spectrometer, several other diagnostics will be used to characterize the proton beams to better understand the physics of self-modulation. Coherent Transition Radiation (in the visible and in the infra-red) produced when the proton beam passes through a thin foil, will be measured using a streak camera. Additionally, transverse coherent transition radiation will be produced and detected using electro-optical sensors; this will be the first experimental use of this recent concept~\cite{pukh12}. First protons to the experiment are expected at the end of 2016 and this will be followed by an initial 3--4 year experimental program of four periods of two weeks of data taking.

\section{Future of the laser plasma accelerators}

The tremendous progress that has been made in plasma acceleration \cite{esar09,malk12,josh07}, from the first acceleration of externally injected electrons in a ${\gvm}$ laser wakefield, self-injection in a $100 {\gvm}$ laser wakefield with first a 100 ${\mev}$ broad spectra to the series of experiments with the production of a quasi-monoenergetic electron beam in a laser wakefield with a compact 10  Hz laser system have contributed to boosting this field of research in which tens of laboratories/teams are playing important roles in a competitive and friendly approach. The evolution of short-pulse laser technology with diode pump lasers or fibre lasers, a field in rapid progress, will eventually contribute to the improvement of laser plasma acceleration and their societal applications, in material science for example for high resolution gamma radiography \cite{glin05, beni11}, in medicine for cancer treatment\cite{glin06b, fuch09b}, in chemistry \cite{broz05, gaud10} and in radiobiology \cite{malk10, riga10, malk08}. In the near future, the development of compact free electron lasers could open the way to the production of intense X-ray beams, in a compact way, by coupling the electron beam with undulators. Thanks to the very high peak current of a few kiloamperes \cite{Lund11}, comparable to the current used at LCLS, the use of laser plasma accelerators for free electron lasers, the so-called fifth generation light source, is clearly identified by the scientific community as a major development. Alternative schemes to produce ultra-short X-ray beams, such as Compton, betatron or Bremsstrahlung X-ray sources, have also been considered. Tremendous progress has been made regarding the study of betatron radiation in a laser plasma accelerator. Since its first observation in 2004 \cite{rous04} and the first monitoring of electron betatronic motion in 2008\cite{glin08}, a number of articles have reported in more detail this new source, including measures of a sub ps duration\cite{taph07} and of a transverse size in the micrometre range \cite{taph06}. Betatron radiation was used recently to perform with high spatial resolution, of about 10 microns, X-ray contrast phase images in a single shot mode operation\cite{four11,knei11}. In parallel, similar huge progress has been performed in accelerating electrons and positrons using electron or positron bunches, with here, a gain of a few tens of ${\gev}$ in a few tens of centimetres accelerating gradient. Wakefields driven by electron beams are good candidates to boost electron energy in a metre long plasma device. The requirement for the driver being very close to the one for FEL purposes, shorter radiation wavelength could be produced by doubling, for example, the electron energy delivered from SLAC or from DESY. Acceleration of electrons and positrons with these drivers are also very relevant for a staging approach for high energy physics purposes.
The AWAKE experiment will certainly contribute to defining the roadmap for future larger-scale R\&D projects on laser, electron or proton driven plasma wakefield acceleration for future high energy colliders for particle physics.
The success of plasma wakefield accelerators will open a pathway towards many exciting societal application, a compact FEL radiation source and a revolutionary plasma-based ${\tev}$ lepton collider. This revolution could then enable ground-breaking discoveries in many domains, including particle physics.

 \section*{Acknowledgements}

I acknowledge the support of the European Research Council for funding the PARIS and X-Five ERC projects (contract number 226424 and 339128), from CARE, EUCARD and UCARD2 and from LASERLAB1,2 and 3. I warmly acknowledge my many excellent collaborators from LOA and all over the world, my new friends R.~Bailey and B.~Holzer from CERN for initiating and organizing with success this first CAS-CERN Accelerator School on Plasma Wake Acceleration.


\end{document}